\documentclass[sigconf, 10pt, screen, dvipsnames]{acmart}
\pdfoutput=1

\makeatletter               
\def\mdseries@tt{m}
\makeatother

\usepackage[english]{babel}
\usepackage{balance}
\usepackage{blindtext}
\usepackage[frozencache,cachedir=.]{minted}
\usepackage{booktabs}
\usepackage{xcolor}
\usepackage{setspace}
\usepackage{tabularx}
\usepackage{csvsimple}
\usepackage{colortbl} 
\usepackage{multirow}
\usepackage{listings}
\usepackage{bm}
\usepackage{microtype}
\usepackage{caption,subcaption}
\usepackage[compact]{titlesec}
\graphicspath{ {./figures/} }

\newcommand*{\belowrulesepcolor}[1]{%
  \noalign{%
    \kern-\belowrulesep
    \begingroup
      \color{#1}%
      \hrule height\belowrulesep
    \endgroup
  }%
}
\newcommand*{\aboverulesepcolor}[1]{%
  \noalign{%
    \begingroup
      \color{#1}%
      \hrule height\aboverulesep
    \endgroup
    \kern-\aboverulesep
  }%
}

\copyrightyear{2019}
\acmYear{2019}
\setcopyright{acmcopyright}
\acmConference[ANRW '19]{ANRW '19: Applied Networking Research Workshop}{July 22, 2019}{Montreal, QC, Canada}
\acmBooktitle{ANRW '19: Applied Networking Research Workshop (ANRW '19), July 22, 2019, Montreal, QC, Canada}
\acmPrice{15.00}
\acmDOI{10.1145/3340301.3341131}
\acmISBN{978-1-4503-6848-3/19/07}

\title{Checking-in on Network Functions}

\author{Zeeshan Lakhani}
\affiliation{%
  \institution{Carnegie Mellon University}
  \city{Pittsburgh}
  \state{Pennsylvania}
}
\email{zlakhani@cs.cmu.edu}

\author{Heather Miller}
\affiliation{%
  \institution{Carnegie Mellon University}
  \city{Pittsburgh}
  \state{Pennsylvania}
}
\email{heather.miller@cs.cmu.edu}

\begin{document}
\sloppy
\begin{abstract}
 When programming network functions, changes within a packet tend to have consequences---side effects which must be accounted for by network programmers or administrators via arbitrary logic and an innate understanding of dependencies. Examples of this include updating checksums when a packet's contents has been modified or adjusting a payload length field of a IPv6 header if another header is added or updated within a packet. While static-typing captures interface specifications and how packet contents should behave, it does not enforce precise invariants around runtime dependencies like the examples above. Instead, during the design phase of network functions, programmers should be given an easier way to specify checks up front, all without having to account for and keep track of these \textit{consequences} at each and every step during the development cycle. In keeping with this view, we present a unique approach for adding and generating both static checks and dynamic contracts for specifying and checking packet processing operations. We develop our technique within an existing framework called NetBricks and demonstrate how our approach simplifies and checks common dependent packet and header processing logic that other systems take for granted, all without adding much overhead during development.
\end{abstract}
\ccsdesc[500]{Networks~Programming interfaces}
\keywords{network functions, design-by-contract, software verification}
\maketitle

\section{Introduction} \label{introduction}
Writing network functions (NFs) today is as capable as ever, with numerous frameworks and domain-specific languages to choose from. Some target development ease, reusable abstractions, or a familiar programming model that is in vogue within software development at large, while others stress performance guarantees or deployment at scale. Irrespective of which framework or model is used, NFs tend to be comprised of code exercising arbitrary logic and domain knowledge that only network programmers or administrators would know, or \textit{should} know.

Consider the payload length field of an IPv6 header, a field whose value is dependent upon the consequence of processing and manipulating the rest of the packet it's a part of. If an extension header \cite {RFC2460} is added to or removed from the packet, or a layer 4 protocol's payload is modified in any way, then this payload length field must be incremented or decremented accordingly. Other ``middleboxes'' downstream in the network will apply functionality based on the value held in this field---without calculating the length of the rest of the packet---or just drop the packet outright if it's wrong. Handling this effect is often taken for granted, a piece of arbitrary logic that network programmers have to remember to apply and validate at different steps in a function pipeline or at egress. For example, in the still widely used \cite {Dobrescu:2009:REP:1629575.1629578, Medhat:2017:SFC:3058315.3058423, Kablan:2015:SNF:2785989.2785993} packet processing system Click \cite {Kohler:2000:CMR:354871.354874}, a
\textit{CheckIP6Header} module provides validation on the payload length field via:

\begin{minted}[fontsize=\small,autogobble,xleftmargin=16pt,linenos,escapeinside=||]{c++}
if(ntohs(ip->ip6_plen) > (plen - |\colorbox{BurntOrange}{40}|))
  |\colorbox{Apricot}{goto bad}|;
\end{minted}

While this code does do a ``check'' on the field, it hard codes the value as part the conditional check instead of using a constant or variable to better express meaning (\colorbox{BurntOrange}{\textit{40}} is the fixed size of an IPv6 header). Additionally, if the check is invalid, \colorbox{Apricot}{\textit{goto bad}} executes a jump, leaving very little in the way of failure handling and unambiguous messaging. Besides a few per-field validations within the element file, the module does not account for related changes downstream when composed with other modules or the possibility of extension headers or variable-length fields. The functionality expressed in this snippet need not be so unwieldy, as validations should be a first-class part of programmable network architecture. 

In this paper, we present a novel approach that clearly describes and validates these arbitrary effects via the addition and generation of both static checks and dynamic, runtime contracts for specifying conditional dependencies in common packet processing actions. Our work makes use of three well-known programming paradigms.

\textbf{\textit{Static Assertions and Types}} Our prototype is incorporated within a framework built using the Rust programming language, which emphasizes a strong, static type system with first-class polymorphism \cite{RustReasoning} (parametric and ad-hoc). Headers within a packet are explicitly-typed, e.g. \textit{Ipv4Hdr} or \textit{Ipv6Hdr} for instance, and contain associated types \cite{rust-book} that define which header(s) can precede it, e.g. an IPv6 Header relies on an Ethernet header existing before it within a packet. We leverage the type system along with the concept of static assertions \cite{StaticAssertions} to provide compile-time checking for a subset of network function components, including constant checks at the call site of these functions and ensuring that packet order and definitions adhere to specification. 

\textbf{\textit{Design by Contract}} We take inspiration from D's contract system \cite{DContracts}, whose design was inspired by contract schemes \cite{Meyer:1992:ADC:618974.619797} where contracts are provided and run during testing, debugging, and development stages---the design phase---but are usually omitted for release builds in order to maximize performance. In using this style of contracts, programmers are able to code, test, and simulate NF's around \textit{pre} (ingress) and \textit{post} (egress) invariants, checking which conditions must hold as packets are transformed by functions. These contracts allow developers to identify the intentional consequences of their packet processing algorithms.

\textbf{\textit{Code Generation}}
Though contract programming aids in checking and asserting if specifications and dependencies between operations hold, these checks are only (recommended to be) provided during time of development or within testing environments. Additionally, we do not want developers to have to sprinkle contracts throughout their NFs or implement logic to traverse or backtrack from one header, payload, or set of bytes to another just for the sake of validation. Instead, our approach allows developers to specify a set of dependencies and conditions up front via macros \cite{Kohlbecker:1987:MDS:41625.41632}, which in turn get translated into contracts.

We develop our technique as an extension to the NetBricks framework and programming model \cite{Panda:2016:NTV:3026877.3026894}, illustrating the efficacy of our approach through two examples: 1) updating the IPv6 payload length of a packet in the context of changes to an extension header; and 2) transitioning an invalid TCP request (based on an MTU---Maximum Transmission Unit---threshold), into that of an ICMPv6 \textit{Packet Too Big} response \cite{RFC4443}. We evaluate our prototype by examining syntax additions, compilation times, and possible runtime overheads. In Section \ref{discussion}, we discuss our examples within the context of a couple real-world implementations, Onos and Facebook network code, where our approach could be beneficial.  

\section{Motivation}
Choosing between NF architectures and/or network programming languages has become a non-trivial process: What kind of programming paradigm should one choose for packet processing, e.g. functional, dataflow, or imperative? Should the framework support the OpenFlow protocol or be composed of its own data plane and control plane layers? What are the most important facets of the system or application: performance, usability and reconfiguration, reliability? There are many choices and abstractions to deliberate on, see sections \ref{discussion} and \ref{introduction}, yet most only provide a subset of safety, design benefits, or degrees of freedom for which kinds of applications can be executed. We illustrate two specific challenges in defining a better way forward. 

\paragraph{\textbf{The Limits of Correctness}} In the interest of handling correctness and ensuring network programs satisfy specification, there are several efforts which have experimented with verifying networking constructs. For example, a language like NetKat \cite{Anderson:2014:NSF:2535838.2535862}, based on proven semantic and type theoretic foundations, provides static checking for reachability, guarantees non-interference between programs, and supports first-class primitives for the filtering, modifying, and transmitting of packets. However, though powerful, NetKat is limited in what logic it can check for and what protocols and actions it can support, as all programs must conform to the OpenFlow flow table---its compilation target. While OpenFlow \textit{is} used in practice, its model is limited in terms of interface, protocol, and field support, especially for newer, experimental features. Due to this coupling with OpenFlow, along with a lack for handling arbitrary logic in packet processing, NetKat does not present a generalized solution.

\paragraph{\textbf{Arbitrary Logic \& Variable-length Data}} \label{var-length} As described in Section \ref{introduction}, many network programs contain operational logic that's only applied based on the IETF or similar specifications they conform to. Some even define their own inspired-by protocols without a formalized spec \cite{Jin2018NetChainSS}. One major component of the IPv6 protocol specification that has been left unsupported by many NF frameworks is that of IPv6 extension headers. Traffic containing such a header is usually dropped in practice and considered a ``threat to the Internet'' \cite{8002912}. In skipping support for extension headers, packet-processing paradigms can avoid dealing with variable-length data---the specs of these headers contain fields with variable-byte-sized data---and complex header chaining dependencies, as these headers can be stacked upon each other to no end. However, as unique applications for programmable networks that make use of these extensions are constantly being explored \cite{7980143}, we must provide programmatic abstractions for adhering to the conditions of these protocols while also being amenable to new, experimental ones down the line whether they're used in industry or proposed in research. 

\section{Kinds of Contracts} \label{kinds}
\subsection{Design by Contract} \label{kinds-doc}
The Eiffel programming language made design by contract first-class, focusing primarily on how runtime contracts can be turned on for monitoring and testing situations so that developers can ``sit back, and just watch their contracts be violated'' \cite{WhyNotProgramRight}.
The key idea behind the approach is that elements of a software system collaborate with each other on the basis of mutual obligations and benefits, driven by dependencies and related components in the system. These contracts are usually separated into \textit{pre} (input/ingress) and \textit{post} conditions (output/egress), where invariants can be asserted on for incoming and outgoing data accordingly. 

In our system, design by contract-styled assertions help programmers articulate what the values of fields in a header should be equal to, bound by, approximate to, or how these values may have shifted during packet transformation (e.g. swapping of MAC addresses). From a processing perspective, the input precondition runs when the packet enters a NF and the postcondition runs as the packet is exiting the function. 

\subsection{Static Assertions}
Static assertions, popularized in the C, C++, and D languages, allow for compile-time assertions of statically defined expressions, e.g. constants, statics. Beyond just checking for specific values, static assertions can be used to enforce fields on \textit{struct} types and check if a pointer's underlying value is the same when coerced to another type. NF programs tend to be comprised of many constants referring to values derived from specifications. For example, the IPv6 minimum MTU value is $1280$ \cite{RFC2460}, but is actually $1294$ in practice when the Ethernet header is included. Our approach can check this caveat statically at the call site where the NF is defined---not where it's instantiated---via compile-time assertions in our prototype for constant checking. Additionally, thanks to \textit{conditional compilation} (see \ref{cond-comp} for more information), static assertions remain in release binaries. 

\subsection{Static Order-Persevering Headers} \label{order-preserve}
With our approach implemented in NetBricks, we were given a head start toward better validation mechanics with a strong, static type system and framework for programming NFs in a map-reduce fashion. To add packet headers in NetBricks, you define a \textit{struct} with the appropriate fields, as you would do in C or P4 for example. All structs must implement a trait\footnote{Traits are used to define shared behavior in Rust, similar to interfaces in other languages \cite{rust-book}.} containing an \textit{associated type} that is defined as \textbf{PreviousHdr}:
\begin{minted}[fontsize=\small,autogobble,xleftmargin=16pt,linenos,escapeinside=||]{rust}
impl EndOffset for Ipv6Hdr {
    |\colorbox{LimeGreen}{type PreviousHdr=EthHdr}|;
    fn offset(&self) -> usize { 40 }
}
\end{minted}

When parsing a packet within an NF, the order is guaranteed by the defined \colorbox{LimeGreen}{\textit{PreviousHdr}}. Given any other order (e.g. parsing an IPv6 header after a ICMPv6 header), the type checker will throw a compile-time error. In our prototype, we leverage this statically-defined order mechanism on headers (\ref {impl-contracts}) to ensure that incoming and outgoing packet header ordering is preserved according to encoded expectations.

\section{Implementation}
We have developed a prototype\footnote{Openly accessible as a \href{https://github.com/williamofockham/NetBricks/tree/zl/contracts-wip}{branch} on Github.} that extends the NetBricks programming model via macro-based metaprogramming with very little additional syntax. Instead of having to manually incorporate or implement all of the contract methodologies described in section \ref{kinds} throughout a NF code base, our contracts extension can be used gradually, i.e. on certain NFs and not others, as well as retroactively on existing NFs with just a few easy steps:
(\textit{i.}) import our \textbf{check} library into an NF module; then
(\textit{ii.}) identify an NF to validate, and mark it; and then (\textit{iii.}) specify contracts at the beginning and end of a NF based on properties that the developer wants to uphold.
 
Once introduced, these contracts rewrite NFs to include mechanisms for storing runtime info (used for checking outgoing packets), generating validations, assertions, logging facilities, and flag checks for conditional compilation.

\paragraph{\textbf{Initialization}} 
As seen in Figure \ref{fig:pre-post}, the \colorbox{Apricot}{\textit{check}} attribute macro (surrounded by brackets) is responsible for three steps in the contract generation process. Firstly, it identifies that the developer wants this NF to be ``checked,'' which means it can be used gradually over time. Secondly, by designating that this function has contract-checking \textbf{on}, we are then able to parse specific keywords, i.e. \textit{pre}, \textit{post}, in the figure that we want to rewrite and generate assertions from. Lastly, it performs a series of initialization operations, including turning-on specialized logging facilities and lazily instantiating a runtime hashmap that's used to store all the headers as part of the input packet to create a mirror of the contents of the packet entering the NF. Building this map allows us to store header information for tracing, analysis, and further checks throughout the processing lifecycle, all the while producing a series of iterative steps to parse through the packet header-by-header, based on the order specified by the code.

\paragraph{\textbf{Macro Expansion}}
The generation of code from macros \textit{ingress\_check!} and \textit{egress\_check!} occurs prior to the NF program being subjected to Rust's type-checker, i.e. occurring in a separate compilation step. Of note, the \colorbox{Lavender}{order} key in both the pre and post assertions, specified by the developer, allows us to match on the header-order within the contents of the packet itself, as all parsing of headers requires explicit type annotations in NetBricks under the hood. If the expected order does not match up on either ingress or egress checks, a compile-time error is thrown (as per \ref{order-preserve}).

\begin{figure}[t!]
\setlength{\belowcaptionskip}{-12pt}
\begin{minted}[escapeinside=||,fontsize=\small]{rust}
#[|\colorbox{Apricot}{check}|(IPV6_MIN_MTU = 1280)]
fn send_too_big {
.pre(box |pkt| {
    ingress_check! {
        input: pkt,
        |\colorbox{Lavender}{order}|: [EthHdr=>Ipv6Hdr=>TcpHdr<Ipv6Hdr>],
        checks: [(|\colorbox{Melon}{payload\_len[Ipv6Hdr]}|, >,
                  |\colorbox{Yellow}{IPV6\_MIN\_MTU}|)]
    }})
...filter/map/group_by operations...
\end{minted}
\begin{minted}[fontsize=\small,escapeinside=||]{rust}
.post(box |pkt| {
    egress_check! {
        input: pkt,
        |\colorbox{Lavender}{order}|:[EthHdr=>Ipv6Hdr=>Icmpv6PktTooBig<...>],
        checks:[(|\colorbox{Melon}{checksum[Icmpv6PktTooBig]}|, neq, 
                 |\colorbox{Yellow}{checksum[TcpHdr<Ipv6Hdr>]}|),
                (|\colorbox{Melon}{payload\_len[Ipv6Hdr]}|, ==, |\colorbox{Yellow}{1240}|),
                (|\colorbox{Melon}{src[Ipv6Hdr]}|, ==, |\colorbox{Yellow}{dst[Ipv6Hdr]}|),
                (|\colorbox{Melon}{dst[Ipv6Hdr]}|, ==, |\colorbox{Yellow}{src[Ipv6Hdr]}|),
                (|\colorbox{Melon}{.src[EthHdr]}|, ==, |\colorbox{Yellow}{.dst[EthHdr]}|),
                (|\colorbox{Melon}{.dst[EthHdr]}|, ==, |\colorbox{Yellow}{.src[EthHdr]}|)]
        }})
\end{minted}
\caption{Pre and Post contracts on MTU example}
\label{fig:pre-post}
\end{figure}

 \paragraph{\textbf{Ingress and Egress Contracts}} \label{impl-contracts}
Figure \ref{fig:pre-post} exhibits how contracts are extended into an NF. This example checks if an incoming TCP packet is beyond the valid MTU threshold, and, if so, then rewrites the packet into that of an ICMPv6 \textit{Packet Too Big} response which gets returned to the source sender. As mentioned, the incoming and outgoing \colorbox{Lavender}{order} lists reveal how the packet should be transformed throughout the main body of the function. Egress checks compare the values of fields and functions on the current, outgoing packet (\colorbox{Melon}{left-hand side of each check}) to values that are either literal integers or integer expressions, or functions or fields from the original, incoming packet (\colorbox{Yellow}{right-hand side}). In this example, if it has to return to sender, this means swapping Ethernet addresses and IPv6 source and destination addresses from the original input. The checks presented here would fail or throw errors if the inner body's logic did not account for these swap operations.

\subsection{Conditional Compilation} \label{cond-comp}
As previously mentioned, design by contract systems were devised with the intention that contracts would be applied during simulation, testing, and debugging stages of development. Our approach combines these kinds of runtime, dynamic assertions, which capture arbitrary logic and values, with static assertions and compile-time type checking. As shown in our evaluation of the runtime cost of our prototype, \ref{runtime-cost}, runtime-checking and initialization accrue a penalty, which is manageable during NF development, not production. We leverage Rust's compile-time feature-flags \cite{FeatureFlags} to only generate dynamic, runtime contracts for debug and testing modes, while ensuring all static information and assertions remain in the finalized, production program binaries.

\section{Evaluation}
In this section, we evaluate the possible overheads of our approach, including profiling its runtime cost by sampling the call graph during a packet's run through an NF. 

\paragraph{\textbf{Setup}}
In our experimental setup, we ran NetBricks within an Ubuntu Docker container on a local VirtualBox VM. NetBricks uses DPDK \cite{Panda:2016:NTV:3026877.3026894} for fast packet I/O, which we have properly set up within the VM and container. We used MoonGen \cite{Emmerich2015MoonGenAS} to generate varying packet captures (pcaps) for our testing and evaluation harness. We looked at three factors in evaluating our technique for the design of NFs: (\textit{i.}) \textbf{additional syntax} (\textit{LoC}---lines of code); (\textit{ii.}) \textbf{compilation-time} added to our two example NFs; (\textit{iii.}) and \textbf{runtime overhead} of ingress and egress contract generation.

\subsection{Syntax Added}
\begin{table}[h!]
\small
\begin{tabular}{@{}lllll@{}}
\toprule
\multicolumn{1}{|l|}{\textbf{LoC run}}          & \multicolumn{1}{l|}{\textbf{lang}} & \multicolumn{1}{l|}{\textbf{files}} & \multicolumn{1}{l|}{\textbf{lines}} & \multicolumn{1}{l|}{\textbf{code}} \\ \midrule
\textit{mtu-too-big: Contracts ON}              & rust                                   & 2                                   & 214                                 & 183                                \\
\textit{mtu-too-big: Contracts OFF}             & rust                                   & 2                                   & 189                                 & 158                                \\
\textit{mtu-too-big: Contracts ON}              & toml                                   & 1                                   & 19                                  & 16                                 \\
\textit{mtu-too-big: Contracts OFF}             & toml                                   & 1                                   & 16                                  & 13                                 \\
\textit{mtu-too-big: Contracts ON}              & total                                  & 3                                   & 233                                 & 199                                \\
\textit{mtu-too-big: Contracts OFF}             & total                                  & 3                                   & 205                                 & 171                                \\ \midrule
\multicolumn{1}{|l|}{\textbf{Change}} & \multicolumn{1}{l|}{}                  & \multicolumn{1}{l|}{0}              & \multicolumn{1}{l|}{+28}             & \multicolumn{1}{l|}{+28}            \\ \bottomrule
\end{tabular}
\caption{Syntax additions for our MTU NF (unexpanded)}
\label{table:LoC}
\end{table}

\begin{table*}[t!]
\small
\begin{tabular}{@{}|l|l|l|l|l|l|l|l|@{}}
\toprule
\textbf{compile times / cargo build} & \textbf{example} & \textbf{mean (s)}  & \textbf{stddev (s)} & \textbf{user (s)}   & system (s)          & min (s)     & max (s)             \\ \midrule
Contracts - Off                      & srv6-change-pkt  & 26.039         & 3.286           & 0.631           & 10.715          & 22.330          & 33.230          \\ \midrule
Contracts - On                       & srv6-change-pkt  & 25.099         & 2.398           & 0.549           & 11.697          & 20.238          & 28.220          \\ \midrule
\belowrulesepcolor{gray!20}
\rowcolor{gray!20}
\textbf{Effect}                      &                  & \cellcolor{green!35}\textbf{-0.94} &  \cellcolor{green!35}\textbf{-0.888} &  \cellcolor{green!35}\textbf{-0.082} &  \cellcolor{red!35}\textbf{+0.982} &  \cellcolor{green!35}\textbf{-2.092} &  \cellcolor{green!35}\textbf{-5.01}  \\ \aboverulesepcolor{gray!20} \midrule
Contracts - Off                      & mtu-too-big      & 21.652         & 2.202           & 0.537           & 9.201           & 18.528          & 25.191          \\
\midrule
Contracts - On                       & mtu-too-big      & 26.052         & 1.858           & 0.650           & 10.851          & 22.165          & 28.346          \\
\midrule
\belowrulesepcolor{gray!20}
\rowcolor{gray!20}
\textbf{Effect}                      &                  & \cellcolor{red!35}\textbf{+4.4}  & \cellcolor{green!35}\textbf{-0.344} & \cellcolor{red!35}\textbf{+0.113} & \cellcolor{red!35}\textbf{+1.65}  & \cellcolor{red!35}\textbf{+3.637} & \cellcolor{red!35}\textbf{+3.155} \\ \aboverulesepcolor{gray!20} \bottomrule
\end{tabular}
\caption{Compile times running ``cargo build'' for extension header NF example and MTU NF example}
\label{table:compile}
\end{table*}

Being that most of the work in our implementation is centered around the macro generation of contracts, it's not surprising to see that our non-expanded measure of \textit{LoC} (Table \ref{table:LoC}) is minimal. We import a few libraries (crates in the Rust ecosystem), including our \textbf{check} library, into NetBricks. The extra crates are used for logging and assertion control around error handling and operations that we can match on. Minus boilerplate, most of the additional code comes from the specifications themselves, as there is no bound on the number of possible validations that can be added.

In Table \ref{table:LoC}, we choose to show \textit{LoC} without expansion, to faithfully represent the experience of the network function developer. At the outset of this project, we wanted to avoid altering many of the core NetBricks APIs or its existing example NFs. With our contract generation prototype, we increased our examples' programs and build configurations by an average of 23 lines, or less than 10 percent.

\subsection{Compilation Times}
One of the most important factors we wanted to consider was compilation time, as we did not want programmers to pay much of a penalty while developing NFs. Table \ref{table:compile} compares our two example network function cases with contract generation turned on and off. For each of these runs, the build was compiled from a warm, incremental cache and then rebuilt from that cache ten times. As shown, with our contract generation system applied, the standard deviation across all builds was less than a second overall. Further, mean time even improved in the case of our extension header example. Though there is room for optimization, these results show that our technique doesn't negatively affect developers during development. 

\subsection{Runtime Cost} \label{runtime-cost}
We've discussed how our objective for programmers is to be able to specify their checks up front as they build out NFs and test them. Knowing all of the initialization and setup we have to concoct on behalf of the contract engine, we were aware that the runtime cost would be problematic if the NF ran in production. But, how problematic would it be?

\begin{figure*}[t!]
\centering
\includegraphics[width=1.0\textwidth]{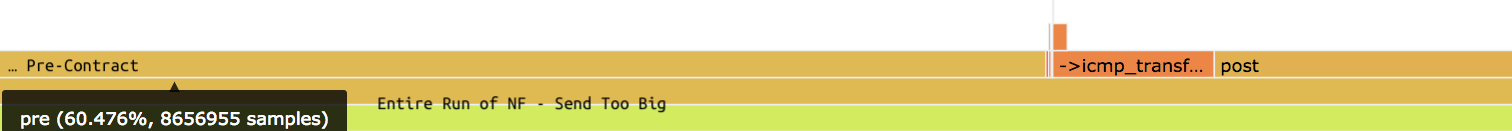}
\caption{Flame graph trace of call time, in samples, for single-packet runs}
\label{fig:flamegraph}
\end{figure*}

For this evaluation, we ran our invalid MTU example and sent a packet at a time through it, tracing the call graph throughout the function and sampling it. To trace and visualize the effect, we used the Flame Graph approach \cite{FlameGraphs}, popularized in industry. The graph is illustrated in Figure \ref{fig:flamegraph}. As expected, our precondition routine takes up a majority of the function's execution time. This is mainly due to creating a copy of the incoming packet and parsing each header within it. Our egress macro, for example, does much less and spans less of the execution graph. 

Further optimizing our implementation's code generation and how we store and parse the packet in the evaluated program would lessen the runtime cost of our technique in practice and eventually make it possible to include some dynamic contracts for in-production use-cases. Nonetheless, as we've stated before, to focus our technique on the design phase of NF writing, the current version of the library compiles away most of this generated code upon production builds---not slowing down the runtime.

\section{Discussion and Future Work}\label{discussion}

Thus far, we've demonstrated our technique on a few simple yet practical NF examples. In this section, we discuss how our work could benefit practical programs and applications out in the wild. Then, we explore where we want to take the approach going forward. 

\subsection{Real-World Example: ONOS Routing} \label{onos}

One of the cases we've evaluated includes adding additional segments to a IPv6 segment routing extension header \cite{I-D.ietf-6man-segment-routing-header}. In Section \ref{var-length}, we mentioned that IPv6 extension headers were difficult to handle due to their variability, causing network operators to write rules to drop packets that contain them and NF frameworks avoiding their logic altogether. ONOS \cite{Berde:2014:OTO:2620728.2620744}, the open network operating system, is a controller platform supporting a wide variety of SDN use cases, including support for the \textit{Routing} header extension. Nonetheless, the most complex logic that the header entails is that of adding and removing segments, which then triggers effects on the \textit{Last Entry} (the index into the stack of segments) and \textit{Segments Left} (the number of route segments remaining to be processed) fields. These triggered events provide a good story for our implementation because the ONOS class \cite{ONUS-ROUTING} does not account for changes in the Routing header stack; instead, it just works on a minimal set of fields for reading these headers along the network path. With an approach like the one we've exhibited in our paper, more assurance could be given for handling the complex logic of variable-length information.  

\subsection{Real World Example: Katran}

Katran \cite{Katran} is Facebook's Layer 4 software load balancer, built on a data plane using the eBPF VM. While Katran is deployed at scale and processes packets at high speeds, its codebase lacks specifications and checks in logic; it also includes many hardcoded values and a myriad of constants. Similar to the invalid MTU example used throughout this paper, they too have a IPv6 \textit{Packet Too Big} response function:

\begin{minted}[xleftmargin=16pt,linenos,escapeinside=||,fontsize=\small]{c++}
static inline int send_icmp6_too_big(...) {
  ...
  icmp6_hdr->icmp6_type = ICMPV6_PKT_TOOBIG;
  icmp6_hdr->icmp6_mtu = htonl(
    |\colorbox{Apricot}{MAX\_PCKT\_SIZE}|-sizeof(struct eth_hdr)
  )}
\end{minted}

While we expect that a test suite would capture possible bugs in arbitrary logic and pointer references, there are no abstractions within the programs themselves to ensure the validity of fields and what values they can possibly be. Furthermore, what if constants like \colorbox{Apricot}{\textit{MAX\_PCKT\_SIZE}} were changed within the upstream module that instantiated them? Leveraging static assertions like we do in our work could benefit functions like the one shown here. 
\subsection{Next Steps}

\subsubsection{\textbf{Deployment Models}}

Throughout this paper, we've explained how runtime assertions in our system are best used during the design phase of programming NFs. Yet within the ecosystem of network programming, our technique can be extended beyond design and integration testing. We would like to be able to \textit{turn on} contracts automatically if our system is running within an environment acting as a simulation network, e.g. Mininet \cite{Lantz2010ANI}, or within production deployments for certain kinds of traffic, e.g. probe packets sent for NF monitoring or health checking.

\subsubsection{\textbf{Hinting and Feedback}}

The major goal of our work is aiding in the development of network programs, especially those that involve arbitrary complexity and interact with changing dependencies. In our system, runtime errors look akin to typical assertion errors presented by other languages or frameworks. Even though our errors include some context---the expectation vs. what actually happened---they normally do not articulate anything related to the specification itself or what dependencies triggered it. We would like to take a page from the recent work in program slicing and compiler design (as demonstrated in languages like Elm and Rust) and provide feedback via hints to the programmer while they build out NFs during development.

\paragraph{\textbf{Leverage static analysis of input programs}}

In our code generation step(s), we look for a set of explicit tokens to rewrite and incorporate seamlessly within the context of a given NF. However, by adding the \textit{check} macro to a function, we're able to walk the entire AST (Abstract Syntax Tree) of the input NF before it gets compiled, allowing us to perform static analysis on the function to find bugs \cite{Hovemeyer:2007:FMN:1251535.1251537} at compile-time. Further leveraging static analysis would allow us to limit the need for certain runtime contracts.

\paragraph{\textbf{Interactive feedback}}

Good feedback is crucial when an error occurs. Modern type systems provide more context to type errors (beyond just which line propagated the error itself), by suggesting more precise types for the developer to include in their programs. In designing network function paradigms, we want to build off our prototype and expand to include helpful information about where contract errors occur at the boundary and in which ways the errors may be debugged. 

\section{Related Work}
Our approach builds upon a growing literature on contract-
\\driven validation of programs. In Sections \ref{introduction} and \ref{kinds-doc} we referenced our inspiration from D's contract programming model, which was itself inspired by the system developed for Eiffel. Though our approach is unique within the field of programmable networks, contract programming has gained popularity as extensions to functional programming languages like Clojure (via Spec \cite{Spec}) and \textit{Racket} \cite{RacketContracts}. Racket also includes mechanisms for generating contracts from macros.

Regarding type systems and ways to validate network programs, we've already mentioned NetKat \cite{Anderson:2014:NSF:2535838.2535862}, which has had follow-up pieces in literature, including probabilistic variants \cite{DBLP:journals/corr/SmolkaKKFK017}. Recently, p4v \cite{Liu:2018:PPV:3230543.3230582} was published and is motivated by real-world examples; it attempts to find bugs in P4 programs and verify program properties by incorporating domain-specific assumptions into a constraint solver.

Languages for network function specification exist within industry, including TOSCA \cite{OASIS2017TOSCA1.0}, a templating \textit{metamodel} for network function virtualization. Also related is \textit{Assert-P4} \cite{Freire:2018:UBP:3185467.3185499}, which is a proposed approach for checking P4 programs that is also based on assertion checking. Combining assertions with symbolic execution, it finds bugs motivated by controller misconfiguration and code circumvention and gives developers that ability to specify properties about their programs. Their work is very P4-specific and does not provide examples of complex pipelines involving arbitrary dependencies, similar to the cases we've discussed in this paper. 

\section{Conclusion}
In this paper, we provide a hybrid-approach and implementation for checking and validating the arbitrary logic and side effects typically part of network functions by combining design by contract, static assertions and type-checking, and code generation via macros. We were able to build-out and incorporate our technique within an existing network function framework, without penalizing the developer or increasing the complexity that they already have to handle. We want to explore this space further and provide better tooling and interaction models for anyone programming networks. 

\clearpage
\section*{Acknowledgments}

We want to thank Professor Justine Sherry from Carnegie Mellon University for inspiring this project while the first author was a part of her Programmable Networks class. Additionally, we're thankful for the guidance and support from Professor Aurojit Panda (New York University) for authoring NetBricks and answering related questions, and Chas Emerick for his all-around help and guidance. Finally, an extra special thank you goes out to the Occam networking team at Comcast (i.e. Peter Cline, Daniel Jin, Andrew Wang, Michael Winslow, Chris Rollins, Paul Cleary, Jon Moore) for their continued support, code, instruction, and feedback. 

\bibliographystyle{ACM-Reference-Format}
\bibliography{reference}


\begin{thebibliography}{33}


\ifx \showCODEN    \undefined \def \showCODEN     #1{\unskip}     \fi
\ifx \showDOI      \undefined \def \showDOI       #1{#1}\fi
\ifx \showISBNx    \undefined \def \showISBNx     #1{\unskip}     \fi
\ifx \showISBNxiii \undefined \def \showISBNxiii  #1{\unskip}     \fi
\ifx \showISSN     \undefined \def \showISSN      #1{\unskip}     \fi
\ifx \showLCCN     \undefined \def \showLCCN      #1{\unskip}     \fi
\ifx \shownote     \undefined \def \shownote      #1{#1}          \fi
\ifx \showarticletitle \undefined \def \showarticletitle #1{#1}   \fi
\ifx \showURL      \undefined \def \showURL       {\relax}        \fi
\providecommand\bibfield[2]{#2}
\providecommand\bibinfo[2]{#2}
\providecommand\natexlab[1]{#1}
\providecommand\showeprint[2][]{arXiv:#2}

\bibitem[\protect\citeauthoryear{??}{ONU}{2017}]%
        {ONUS-ROUTING}
 \bibinfo{year}{2017}\natexlab{}.
\newblock \bibinfo{title}{Routing.java}.
\newblock
  \bibinfo{howpublished}{\url{https://github.com/opennetworkinglab/onos/blob/021d2eb175b8e46d4690cd9e1243301ddd903bcc/utils/misc/src/main/java/org/onlab/packet/ipv6/Routing.java}}.
    (\bibinfo{year}{2017}).
\newblock


\bibitem[\protect\citeauthoryear{??}{Rac}{2018}]%
        {RacketContracts}
 \bibinfo{year}{2018}\natexlab{}.
\newblock \bibinfo{title}{Racket Contracts}.
\newblock   (\bibinfo{year}{2018}).
\newblock
\showURL{%
\url{https://docs.racket-lang.org/guide/contracts.html}}


\bibitem[\protect\citeauthoryear{Anderson, Foster, Guha, Jeannin, Kozen,
  Schlesinger, and Walker}{Anderson et~al\mbox{.}}{2014}]%
        {Anderson:2014:NSF:2535838.2535862}
\bibfield{author}{\bibinfo{person}{Carolyn~Jane Anderson},
  \bibinfo{person}{Nate Foster}, \bibinfo{person}{Arjun Guha},
  \bibinfo{person}{Jean-Baptiste Jeannin}, \bibinfo{person}{Dexter Kozen},
  \bibinfo{person}{Cole Schlesinger}, {and} \bibinfo{person}{David Walker}.}
  \bibinfo{year}{2014}\natexlab{}.
\newblock \showarticletitle{NetKAT: Semantic Foundations for Networks}. In
  \bibinfo{booktitle}{{\em Proceedings of the 41st ACM SIGPLAN-SIGACT Symposium
  on Principles of Programming Languages}} {\em (\bibinfo{series}{POPL '14})}.
  \bibinfo{publisher}{ACM}, \bibinfo{address}{New York, NY, USA},
  \bibinfo{pages}{113--126}.
\newblock
\showISBNx{978-1-4503-2544-8}
\showDOI{%
\url{https://doi.org/10.1145/2535838.2535862}}


\bibitem[\protect\citeauthoryear{Berde, Gerola, Hart, Higuchi, Kobayashi,
  Koide, Lantz, O'Connor, Radoslavov, Snow, and Parulkar}{Berde
  et~al\mbox{.}}{2014}]%
        {Berde:2014:OTO:2620728.2620744}
\bibfield{author}{\bibinfo{person}{Pankaj Berde}, \bibinfo{person}{Matteo
  Gerola}, \bibinfo{person}{Jonathan Hart}, \bibinfo{person}{Yuta Higuchi},
  \bibinfo{person}{Masayoshi Kobayashi}, \bibinfo{person}{Toshio Koide},
  \bibinfo{person}{Bob Lantz}, \bibinfo{person}{Brian O'Connor},
  \bibinfo{person}{Pavlin Radoslavov}, \bibinfo{person}{William Snow}, {and}
  \bibinfo{person}{Guru Parulkar}.} \bibinfo{year}{2014}\natexlab{}.
\newblock \showarticletitle{ONOS: Towards an Open, Distributed SDN OS}. In
  \bibinfo{booktitle}{{\em Proceedings of the Third Workshop on Hot Topics in
  Software Defined Networking}} {\em (\bibinfo{series}{HotSDN '14})}.
  \bibinfo{publisher}{ACM}, \bibinfo{address}{New York, NY, USA},
  \bibinfo{pages}{1--6}.
\newblock
\showISBNx{978-1-4503-2989-7}
\showDOI{%
\url{https://doi.org/10.1145/2620728.2620744}}


\bibitem[\protect\citeauthoryear{Conta, Deering, and Gupta}{Conta
  et~al\mbox{.}}{2006}]%
        {RFC4443}
\bibfield{author}{\bibinfo{person}{A. Conta}, \bibinfo{person}{S. Deering},
  {and} \bibinfo{person}{M. Gupta}.} \bibinfo{year}{2006}\natexlab{}.
\newblock \bibinfo{booktitle}{{\em Internet Control Message Protocol (ICMPv6)
  for the Internet Protocol Version 6 (IPv6) Specification}}.
\newblock \bibinfo{type}{RFC} 4443. \bibinfo{institution}{RFC Editor}.
\newblock
\showISSN{2070-1721}
\showURL{%
\url{http://www.rfc-editor.org/rfc/rfc4443.txt}}
\newblock
\shownote{\url{http://www.rfc-editor.org/rfc/rfc4443.txt}.}


\bibitem[\protect\citeauthoryear{Deering and Hinden}{Deering and
  Hinden}{1998}]%
        {RFC2460}
\bibfield{author}{\bibinfo{person}{Stephen~E. Deering} {and}
  \bibinfo{person}{Robert~M. Hinden}.} \bibinfo{year}{1998}\natexlab{}.
\newblock \bibinfo{booktitle}{{\em Internet Protocol, Version 6 (IPv6)
  Specification}}.
\newblock \bibinfo{type}{RFC} 2460. \bibinfo{institution}{RFC Editor}.
\newblock
\showISSN{2070-1721}
\showURL{%
\url{http://www.rfc-editor.org/rfc/rfc2460.txt}}
\newblock
\shownote{\url{http://www.rfc-editor.org/rfc/rfc2460.txt}.}


\bibitem[\protect\citeauthoryear{Desmouceaux, Pfister, Tollet, Townsley, and
  Clausen}{Desmouceaux et~al\mbox{.}}{2017}]%
        {7980143}
\bibfield{author}{\bibinfo{person}{Y. Desmouceaux}, \bibinfo{person}{P.
  Pfister}, \bibinfo{person}{J. Tollet}, \bibinfo{person}{M. Townsley}, {and}
  \bibinfo{person}{T. Clausen}.} \bibinfo{year}{2017}\natexlab{}.
\newblock \showarticletitle{SRLB: The Power of Choices in Load Balancing with
  Segment Routing}. In \bibinfo{booktitle}{{\em 2017 IEEE 37th International
  Conference on Distributed Computing Systems (ICDCS)}}.
  \bibinfo{pages}{2011--2016}.
\newblock
\showISSN{1063-6927}
\showDOI{%
\url{https://doi.org/10.1109/ICDCS.2017.180}}


\bibitem[\protect\citeauthoryear{Developers}{Developers}{2018}]%
        {rust-book}
\bibfield{author}{\bibinfo{person}{The Rust~Project Developers}.}
  \bibinfo{year}{2018}\natexlab{}.
\newblock \bibinfo{title}{The {Rust} Programming Language, 2st ed}.
\newblock
  \bibinfo{howpublished}{\url{https://doc.rust-lang.org/1.30.0/book/2018-edition/}}.
    (\bibinfo{year}{2018}).
\newblock


\bibitem[\protect\citeauthoryear{Dobrescu, Egi, Argyraki, Chun, Fall,
  Iannaccone, Knies, Manesh, and Ratnasamy}{Dobrescu et~al\mbox{.}}{2009}]%
        {Dobrescu:2009:REP:1629575.1629578}
\bibfield{author}{\bibinfo{person}{Mihai Dobrescu}, \bibinfo{person}{Norbert
  Egi}, \bibinfo{person}{Katerina Argyraki}, \bibinfo{person}{Byung-Gon Chun},
  \bibinfo{person}{Kevin Fall}, \bibinfo{person}{Gianluca Iannaccone},
  \bibinfo{person}{Allan Knies}, \bibinfo{person}{Maziar Manesh}, {and}
  \bibinfo{person}{Sylvia Ratnasamy}.} \bibinfo{year}{2009}\natexlab{}.
\newblock \showarticletitle{RouteBricks: Exploiting Parallelism to Scale
  Software Routers}. In \bibinfo{booktitle}{{\em Proceedings of the ACM SIGOPS
  22Nd Symposium on Operating Systems Principles}} {\em (\bibinfo{series}{SOSP
  '09})}. \bibinfo{publisher}{ACM}, \bibinfo{address}{New York, NY, USA},
  \bibinfo{pages}{15--28}.
\newblock
\showISBNx{978-1-60558-752-3}
\showDOI{%
\url{https://doi.org/10.1145/1629575.1629578}}


\bibitem[\protect\citeauthoryear{Emmerich, Gallenm{\"u}ller, Raumer, Wohlfart,
  and Carle}{Emmerich et~al\mbox{.}}{2015}]%
        {Emmerich2015MoonGenAS}
\bibfield{author}{\bibinfo{person}{Paul Emmerich}, \bibinfo{person}{Sebastian
  Gallenm{\"u}ller}, \bibinfo{person}{Daniel Raumer}, \bibinfo{person}{Florian
  Wohlfart}, {and} \bibinfo{person}{Georg Carle}.}
  \bibinfo{year}{2015}\natexlab{}.
\newblock \showarticletitle{MoonGen: A Scriptable High-Speed Packet Generator}.
  In \bibinfo{booktitle}{{\em Internet Measurement Conference}}.
\newblock


\bibitem[\protect\citeauthoryear{Filsfils, Previdi, Leddy, Matsushima, and
  daniel.voyer@bell.ca}{Filsfils et~al\mbox{.}}{2018}]%
        {I-D.ietf-6man-segment-routing-header}
\bibfield{author}{\bibinfo{person}{Clarence Filsfils}, \bibinfo{person}{Stefano
  Previdi}, \bibinfo{person}{John Leddy}, \bibinfo{person}{Satoru Matsushima},
  {and} \bibinfo{person}{daniel.voyer@bell.ca}.}
  \bibinfo{year}{2018}\natexlab{}.
\newblock \bibinfo{booktitle}{{\em IPv6 Segment Routing Header (SRH)}}.
\newblock \bibinfo{type}{Internet-Draft}
  draft-ietf-6man-segment-routing-header-15. \bibinfo{institution}{IETF
  Secretariat}.
\newblock
\showURL{%
\url{http://www.ietf.org/internet-drafts/draft-ietf-6man-segment-routing-header-15.txt}}
\newblock
\shownote{\url{http://www.ietf.org/internet-drafts/draft-ietf-6man-segment-routing-header-15.txt}.}


\bibitem[\protect\citeauthoryear{Foundation}{Foundation}{2018}]%
        {DContracts}
\bibfield{author}{\bibinfo{person}{D~Language Foundation}.}
  \bibinfo{year}{2018}\natexlab{}.
\newblock \bibinfo{title}{Contract Programming}.
\newblock   (\bibinfo{year}{2018}).
\newblock
\showURL{%
\url{https://dlang.org/spec/contracts.html}}


\bibitem[\protect\citeauthoryear{Freire, Neves, Leal, Levchenko,
  Schaeffer-Filho, and Barcellos}{Freire et~al\mbox{.}}{2018}]%
        {Freire:2018:UBP:3185467.3185499}
\bibfield{author}{\bibinfo{person}{Lucas Freire}, \bibinfo{person}{Miguel
  Neves}, \bibinfo{person}{Lucas Leal}, \bibinfo{person}{Kirill Levchenko},
  \bibinfo{person}{Alberto Schaeffer-Filho}, {and} \bibinfo{person}{Marinho
  Barcellos}.} \bibinfo{year}{2018}\natexlab{}.
\newblock \showarticletitle{Uncovering Bugs in P4 Programs with Assertion-based
  Verification}. In \bibinfo{booktitle}{{\em Proceedings of the Symposium on
  SDN Research}} {\em (\bibinfo{series}{SOSR '18})}. \bibinfo{publisher}{ACM},
  \bibinfo{address}{New York, NY, USA}, Article \bibinfo{articleno}{4},
  \bibinfo{numpages}{7}~pages.
\newblock
\showISBNx{978-1-4503-5664-0}
\showDOI{%
\url{https://doi.org/10.1145/3185467.3185499}}


\bibitem[\protect\citeauthoryear{Gregg}{Gregg}{2018}]%
        {FlameGraphs}
\bibfield{author}{\bibinfo{person}{Brendan Gregg}.}
  \bibinfo{year}{2018}\natexlab{}.
\newblock \bibinfo{title}{Flame Graphs}.
\newblock   (\bibinfo{year}{2018}).
\newblock
\showURL{%
\url{http://www.brendangregg.com/flamegraphs.html}}


\bibitem[\protect\citeauthoryear{Hendriks, Velan, d.~O.~Schmidt, de~Boer, and
  Pras}{Hendriks et~al\mbox{.}}{2017}]%
        {8002912}
\bibfield{author}{\bibinfo{person}{L. Hendriks}, \bibinfo{person}{P. Velan},
  \bibinfo{person}{R. d. O.~Schmidt}, \bibinfo{person}{P. de Boer}, {and}
  \bibinfo{person}{A. Pras}.} \bibinfo{year}{2017}\natexlab{}.
\newblock \showarticletitle{Threats and surprises behind IPv6 extension
  headers}. In \bibinfo{booktitle}{{\em 2017 Network Traffic Measurement and
  Analysis Conference (TMA)}}. \bibinfo{pages}{1--9}.
\newblock
\showDOI{%
\url{https://doi.org/10.23919/TMA.2017.8002912}}


\bibitem[\protect\citeauthoryear{Hovemeyer and Pugh}{Hovemeyer and
  Pugh}{2007}]%
        {Hovemeyer:2007:FMN:1251535.1251537}
\bibfield{author}{\bibinfo{person}{David Hovemeyer} {and}
  \bibinfo{person}{William Pugh}.} \bibinfo{year}{2007}\natexlab{}.
\newblock \showarticletitle{Finding More Null Pointer Bugs, but Not Too Many}.
  In \bibinfo{booktitle}{{\em Proceedings of the 7th ACM SIGPLAN-SIGSOFT
  Workshop on Program Analysis for Software Tools and Engineering}} {\em
  (\bibinfo{series}{PASTE '07})}. \bibinfo{publisher}{ACM},
  \bibinfo{address}{New York, NY, USA}, \bibinfo{pages}{9--14}.
\newblock
\showISBNx{978-1-59593-595-3}
\showDOI{%
\url{https://doi.org/10.1145/1251535.1251537}}


\bibitem[\protect\citeauthoryear{Jin, Li, Zhang, Foster, Lee, Soul{\'e}, Kim,
  and Stoica}{Jin et~al\mbox{.}}{2018}]%
        {Jin2018NetChainSS}
\bibfield{author}{\bibinfo{person}{Xin Jin}, \bibinfo{person}{Xiaozhou Li},
  \bibinfo{person}{Haoyu Zhang}, \bibinfo{person}{Nate Foster},
  \bibinfo{person}{Jeongkeun Lee}, \bibinfo{person}{Robert Soul{\'e}},
  \bibinfo{person}{Changhoon Kim}, {and} \bibinfo{person}{Ion Stoica}.}
  \bibinfo{year}{2018}\natexlab{}.
\newblock \showarticletitle{NetChain: Scale-Free Sub-RTT Coordination}. In
  \bibinfo{booktitle}{{\em NSDI}}.
\newblock


\bibitem[\protect\citeauthoryear{Kablan, Caldwell, Han, Jamjoom, and
  Keller}{Kablan et~al\mbox{.}}{2015}]%
        {Kablan:2015:SNF:2785989.2785993}
\bibfield{author}{\bibinfo{person}{Murad Kablan}, \bibinfo{person}{Blake
  Caldwell}, \bibinfo{person}{Richard Han}, \bibinfo{person}{Hani Jamjoom},
  {and} \bibinfo{person}{Eric Keller}.} \bibinfo{year}{2015}\natexlab{}.
\newblock \showarticletitle{Stateless Network Functions}. In
  \bibinfo{booktitle}{{\em Proceedings of the 2015 ACM SIGCOMM Workshop on Hot
  Topics in Middleboxes and Network Function Virtualization}} {\em
  (\bibinfo{series}{HotMiddlebox '15})}. \bibinfo{publisher}{ACM},
  \bibinfo{address}{New York, NY, USA}, \bibinfo{pages}{49--54}.
\newblock
\showISBNx{978-1-4503-3540-9}
\showDOI{%
\url{https://doi.org/10.1145/2785989.2785993}}


\bibitem[\protect\citeauthoryear{Klarer, Maddock, Dawes, and Hinnant}{Klarer
  et~al\mbox{.}}{2008}]%
        {StaticAssertions}
\bibfield{author}{\bibinfo{person}{Robert Klarer}, \bibinfo{person}{John
  Maddock}, \bibinfo{person}{Beman Dawes}, {and} \bibinfo{person}{Howard
  Hinnant}.} \bibinfo{year}{2008}\natexlab{}.
\newblock \bibinfo{title}{Static Assertions}.
\newblock
  \bibinfo{howpublished}{\url{http://www.open-std.org/jtc1/sc22/wg14/www/docs/n1330.pdf}}.
    (\bibinfo{year}{2008}).
\newblock


\bibitem[\protect\citeauthoryear{Kohlbecker and Wand}{Kohlbecker and
  Wand}{1987}]%
        {Kohlbecker:1987:MDS:41625.41632}
\bibfield{author}{\bibinfo{person}{E.~E. Kohlbecker} {and} \bibinfo{person}{M.
  Wand}.} \bibinfo{year}{1987}\natexlab{}.
\newblock \showarticletitle{Macro-by-example: Deriving Syntactic
  Transformations from Their Specifications}. In \bibinfo{booktitle}{{\em
  Proceedings of the 14th ACM SIGACT-SIGPLAN Symposium on Principles of
  Programming Languages}} {\em (\bibinfo{series}{POPL '87})}.
  \bibinfo{publisher}{ACM}, \bibinfo{address}{New York, NY, USA},
  \bibinfo{pages}{77--84}.
\newblock
\showISBNx{0-89791-215-2}
\showDOI{%
\url{https://doi.org/10.1145/41625.41632}}


\bibitem[\protect\citeauthoryear{Kohler, Morris, Chen, Jannotti, and
  Kaashoek}{Kohler et~al\mbox{.}}{2000}]%
        {Kohler:2000:CMR:354871.354874}
\bibfield{author}{\bibinfo{person}{Eddie Kohler}, \bibinfo{person}{Robert
  Morris}, \bibinfo{person}{Benjie Chen}, \bibinfo{person}{John Jannotti},
  {and} \bibinfo{person}{M.~Frans Kaashoek}.} \bibinfo{year}{2000}\natexlab{}.
\newblock \showarticletitle{The Click Modular Router}.
\newblock \bibinfo{journal}{{\em ACM Trans. Comput. Syst.\/}}
  \bibinfo{volume}{18}, \bibinfo{number}{3} (\bibinfo{date}{Aug.}
  \bibinfo{year}{2000}), \bibinfo{pages}{263--297}.
\newblock
\showISSN{0734-2071}
\showDOI{%
\url{https://doi.org/10.1145/354871.354874}}


\bibitem[\protect\citeauthoryear{Lantz, Heller, and McKeown}{Lantz
  et~al\mbox{.}}{2010}]%
        {Lantz2010ANI}
\bibfield{author}{\bibinfo{person}{Bob Lantz}, \bibinfo{person}{Brandon
  Heller}, {and} \bibinfo{person}{Nick McKeown}.}
  \bibinfo{year}{2010}\natexlab{}.
\newblock \showarticletitle{A network in a laptop: rapid prototyping for
  software-defined networks}. In \bibinfo{booktitle}{{\em HotNets}}.
\newblock


\bibitem[\protect\citeauthoryear{Liu, Hallahan, Schlesinger, Sharif, Lee,
  Soul{\'e}, Wang, Ca\c{s}caval, McKeown, and Foster}{Liu
  et~al\mbox{.}}{2018}]%
        {Liu:2018:PPV:3230543.3230582}
\bibfield{author}{\bibinfo{person}{Jed Liu}, \bibinfo{person}{William
  Hallahan}, \bibinfo{person}{Cole Schlesinger}, \bibinfo{person}{Milad
  Sharif}, \bibinfo{person}{Jeongkeun Lee}, \bibinfo{person}{Robert Soul{\'e}},
  \bibinfo{person}{Han Wang}, \bibinfo{person}{C\u{a}lin Ca\c{s}caval},
  \bibinfo{person}{Nick McKeown}, {and} \bibinfo{person}{Nate Foster}.}
  \bibinfo{year}{2018}\natexlab{}.
\newblock \showarticletitle{P4V: Practical Verification for Programmable Data
  Planes}. In \bibinfo{booktitle}{{\em Proceedings of the 2018 Conference of
  the ACM Special Interest Group on Data Communication}} {\em
  (\bibinfo{series}{SIGCOMM '18})}. \bibinfo{publisher}{ACM},
  \bibinfo{address}{New York, NY, USA}, \bibinfo{pages}{490--503}.
\newblock
\showISBNx{978-1-4503-5567-4}
\showDOI{%
\url{https://doi.org/10.1145/3230543.3230582}}


\bibitem[\protect\citeauthoryear{Medhat, Taleb, Elmangoush, Carella, Covaci,
  and Magedanz}{Medhat et~al\mbox{.}}{2017}]%
        {Medhat:2017:SFC:3058315.3058423}
\bibfield{author}{\bibinfo{person}{Ahmed~M. Medhat}, \bibinfo{person}{Tarik
  Taleb}, \bibinfo{person}{Asma Elmangoush}, \bibinfo{person}{Giuseppe~A.
  Carella}, \bibinfo{person}{Stefan Covaci}, {and} \bibinfo{person}{Thomas
  Magedanz}.} \bibinfo{year}{2017}\natexlab{}.
\newblock \showarticletitle{Service Function Chaining in Next Generation
  Networks: State of the Art and Research Challenges}.
\newblock \bibinfo{journal}{{\em Comm. Mag.\/}} \bibinfo{volume}{55},
  \bibinfo{number}{2} (\bibinfo{date}{Feb.} \bibinfo{year}{2017}),
  \bibinfo{pages}{216--223}.
\newblock
\showISSN{0163-6804}
\showDOI{%
\url{https://doi.org/10.1109/MCOM.2016.1600219RP}}


\bibitem[\protect\citeauthoryear{Meyer}{Meyer}{1992}]%
        {Meyer:1992:ADC:618974.619797}
\bibfield{author}{\bibinfo{person}{Bertrand Meyer}.}
  \bibinfo{year}{1992}\natexlab{}.
\newblock \showarticletitle{Applying "Design by Contract"}.
\newblock \bibinfo{journal}{{\em Computer\/}} \bibinfo{volume}{25},
  \bibinfo{number}{10} (\bibinfo{date}{Oct.} \bibinfo{year}{1992}),
  \bibinfo{pages}{40--51}.
\newblock
\showISSN{0018-9162}
\showDOI{%
\url{https://doi.org/10.1109/2.161279}}


\bibitem[\protect\citeauthoryear{Meyer}{Meyer}{2018}]%
        {WhyNotProgramRight}
\bibfield{author}{\bibinfo{person}{Bertrand Meyer}.}
  \bibinfo{year}{2018}\natexlab{}.
\newblock \bibinfo{title}{Why not program right?}
\newblock   (\bibinfo{year}{2018}).
\newblock
\showURL{%
\url{https://bertrandmeyer.com/2018/05/24/not-program-right}}


\bibitem[\protect\citeauthoryear{Miller}{Miller}{2018}]%
        {Spec}
\bibfield{author}{\bibinfo{person}{Alex Miller}.}
  \bibinfo{year}{2018}\natexlab{}.
\newblock \bibinfo{title}{spec Guide}.
\newblock   (\bibinfo{year}{2018}).
\newblock
\showURL{%
\url{https://clojure.org/about/spec}}


\bibitem[\protect\citeauthoryear{{Organization for the Advancement of
  Structured Information Standards}}{{Organization for the Advancement of
  Structured Information Standards}}{2017}]%
        {OASIS2017TOSCA1.0}
\bibfield{author}{\bibinfo{person}{{Organization for the Advancement of
  Structured Information Standards}}.} \bibinfo{year}{2017}\natexlab{}.
\newblock \bibinfo{title}{{TOSCA Simple Profile for Network Functions
  Virtualization (NFV)—Version 1.0}}.
\newblock   (\bibinfo{year}{2017}).
\newblock
\showURL{%
\url{http://docs.oasis-open.org/tosca/tosca-nfv/v1.0/tosca-nfv-v1.0.html}}


\bibitem[\protect\citeauthoryear{Panda, Han, Jang, Walls, Ratnasamy, and
  Shenker}{Panda et~al\mbox{.}}{2016}]%
        {Panda:2016:NTV:3026877.3026894}
\bibfield{author}{\bibinfo{person}{Aurojit Panda}, \bibinfo{person}{Sangjin
  Han}, \bibinfo{person}{Keon Jang}, \bibinfo{person}{Melvin Walls},
  \bibinfo{person}{Sylvia Ratnasamy}, {and} \bibinfo{person}{Scott Shenker}.}
  \bibinfo{year}{2016}\natexlab{}.
\newblock \showarticletitle{NetBricks: Taking the V out of NFV}. In
  \bibinfo{booktitle}{{\em Proceedings of the 12th USENIX Conference on
  Operating Systems Design and Implementation}} {\em
  (\bibinfo{series}{OSDI'16})}. \bibinfo{publisher}{USENIX Association},
  \bibinfo{address}{Berkeley, CA, USA}, \bibinfo{pages}{203--216}.
\newblock
\showISBNx{978-1-931971-33-1}
\showURL{%
\url{http://dl.acm.org/citation.cfm?id=3026877.3026894}}


\bibitem[\protect\citeauthoryear{Shirokov and Dasineni}{Shirokov and
  Dasineni}{2018}]%
        {Katran}
\bibfield{author}{\bibinfo{person}{Nikita Shirokov} {and}
  \bibinfo{person}{Ranjeeth Dasineni}.} \bibinfo{year}{2018}\natexlab{}.
\newblock \bibinfo{title}{Open-sourcing Katran, a scalable network load
  balancer}.
\newblock   (\bibinfo{year}{2018}).
\newblock
\showURL{%
\url{https://code.fb.com/open-source/open-sourcing-katran-a-scalable-network-load-balancer/}}


\bibitem[\protect\citeauthoryear{Smolka, Kahn, Kumar, Foster, Kozen, and
  Silva}{Smolka et~al\mbox{.}}{2017}]%
        {DBLP:journals/corr/SmolkaKKFK017}
\bibfield{author}{\bibinfo{person}{Steffen Smolka}, \bibinfo{person}{David
  Kahn}, \bibinfo{person}{Praveen Kumar}, \bibinfo{person}{Nate Foster},
  \bibinfo{person}{Dexter Kozen}, {and} \bibinfo{person}{Alexandra Silva}.}
  \bibinfo{year}{2017}\natexlab{}.
\newblock \showarticletitle{Deciding Probabilistic Program Equivalence in
  NetKAT}.
\newblock \bibinfo{journal}{{\em CoRR\/}}  \bibinfo{volume}{abs/1707.02772}
  (\bibinfo{year}{2017}).
\newblock
\showeprint[arxiv]{1707.02772}
\showURL{%
\url{http://arxiv.org/abs/1707.02772}}


\bibitem[\protect\citeauthoryear{Weiss}{Weiss}{2018}]%
        {RustReasoning}
\bibfield{author}{\bibinfo{person}{Aaron Weiss}.}
  \bibinfo{year}{2018}\natexlab{}.
\newblock \bibinfo{title}{Reasoning with Types in Rust}.
\newblock   (\bibinfo{year}{2018}).
\newblock
\showURL{%
\url{https://aaronweiss.us/posts/2018-02-26-reasoning-with-types-in-rust.html}}


\bibitem[\protect\citeauthoryear{Worthe}{Worthe}{2018}]%
        {FeatureFlags}
\bibfield{author}{\bibinfo{person}{Justin Worthe}.}
  \bibinfo{year}{2018}\natexlab{}.
\newblock \bibinfo{title}{Compile Time Feature Flags in Rust}.
\newblock   (\bibinfo{year}{2018}).
\newblock
\showURL{%
\url{https://www.worthe-it.co.za/programming/2018/11/18/compile-time-feature-flags-in-rust.html}}


\end{thebibliography}
\end{document}